\documentclass[referee]{aa}

\usepackage{graphicx}
\usepackage{natbib}
\usepackage{latexsym}
\usepackage{amsmath}
\usepackage{amssymb}
\usepackage{amstext}
\usepackage{color}
\usepackage{psfrag}
\usepackage{txfonts}

\def\k{km s$^{-1}$}
\def\ks{km s$^{-1}$~}
\def\d{$^\circ$}
\def\m{$^\prime$}
\def\s{$^{\prime\prime}$}
\def\hh{$^{\mathrm h}$}
\def\mm{$^{\mathrm m}$}
\def\ss{$^{\mathrm s}$}
\def\cm3{cm$^{-3}$}

\def\2{$^{12}$CO}
\def\3{$^{13}$CO}
\def\msol{$M_\odot$}

\begin{document}

\title{Study of the luminous blue variable star candidate G26.47+0.02 and its environment }
\author {S. Paron \inst{1,2,3}
\and J. A. Combi \inst{4,5}
\and A. Petriella \inst{1,3}
\and E. Giacani \inst{1,2}
}

\institute{Instituto de Astronom\'\i a y F\'\i sica del Espacio (IAFE),
             CC 67, Suc. 28, 1428 Buenos Aires, Argentina\\
             \email{sparon@iafe.uba.ar}
\and FADU - Universidad de Buenos Aires, Ciudad Universitaria, Buenos Aires, Argentina
\and CBC - Universidad de Buenos Aires, Ciudad Universitaria, Buenos Aires, Argentina
\and Facultad de Ciencias Astron\'omicas y Geof\'{\i}sicas, Universidad Nacional de La Plata, Paseo del Bosque, 1900FWA La Plata, Argentina
\and IAR, CONICET, CCT La Plata, C.C. No. 5 (1894) Villa Elisa, Buenos Aires, Argentina
}

\offprints{S. Paron}

   \date{Received <date>; Accepted <date>}

\abstract{}{The luminous blue variable (LBV) stars are peculiar very massive stars. The study of these stellar objects 
and their surroundings is important for understanding the evolution of massive stars and its effects on the interstellar medium. We 
study the LBV star candidate G26.47+0.02.}
{Using several large-scale surveys in different frequencies we performed a multiwavelength study of G26.47+0.02 and its surroundings.}
{We found a molecular shell (seen in the $^{13}$CO J=1--0 line) that partially surrounds the mid-infrared nebula of G26.47+0.02, which
suggests an interaction between the strong stellar winds 
and the molecular gas. From the HI absorption and the molecular gas study we conclude that G26.47+0.02 is located at a distance 
of $\sim 4.8$ kpc. The radio continuum analysis shows a both thermal and non-thermal emission toward this LBV candidate, 
pointing to wind-wind collision shocks from a binary system. This hypothesis
is supported by a search of near-IR sources and the Chandra X-ray analysis. Additional multiwavelength and long-term
observations are needed to detect some possible variable behavior, and if that is found, to confirm the binary nature of the system.}{}

\titlerunning{The LBV candidate star G26.47+0.02 }
\authorrunning{S. Paron et al.}


\maketitle

\section{Introduction}

Luminous blue variable (LBV) stars are peculiar very massive stars that evolve from the O-type main sequence burning hydrogen 
in their core to become a Wolf-Rayet (WR) helium core burning star. Their main characteristics are a high mass-loss rate 
(up to $10^{-4}$ \msol yr$^{-1}$), 
sometimes accompanied by so-called giant eruptions, a high luminosity ($\sim 10^{6}$ $L_{\odot}$), and significant 
photometric as well as spectroscopic variability \citep{hump94}. The high mass-loss rate associated with the LBV phase 
typically results in the formation
of an ejecta nebula around the star (e.g. \citealt{nota95,clark03}). The nebulae around LBV stars generally are strong 
emitters in the mid-infrared, showing that they are composed of both gaseous and dusty components. 

At present, there are very few cases in which molecular material related to LBV or LBV-candidate nebulae have been determined 
by molecular studies. The most representative 
sources are AG Car \citep{nota02} and G79.29+0.46 \citep{rizzo08,jimenez10}. Very recently, \citet{albert11} discovered 
a fragmented molecular shell delineating the infrared bipolar outer shell of the LBV star G24.73+0.69. The authors argued 
that the molecular shell formed from the interstellar material that was swept-up by the 
stellar wind of the central star. On the other hand, they detected some molecular emission probably associated with the 
inner G24.73+0.69 infrared nebula, which may have originated in a mass-ejection event from the central star.

\begin{figure}[h]
\centering
\includegraphics[width=8cm]{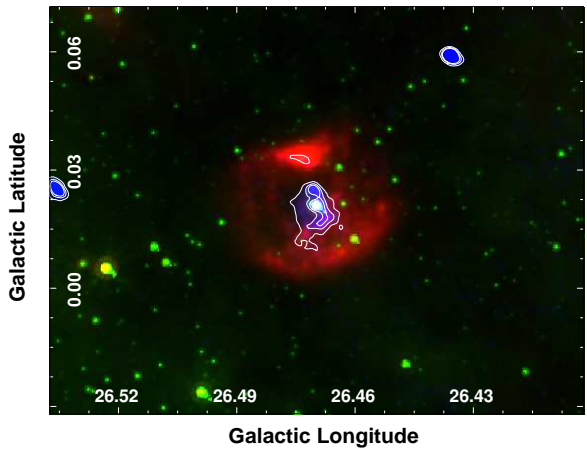}
\caption{Color-composite image of the G26 surroundings, where the {\it Spitzer}-IRAC 8 $\mu$m emission is displayed in green, 
the {\it Spitzer}-MIPSGAL emission at 24 $\mu$m in red, and the radio continuum emission at 20 cm is presented in blue with 
white contours with levels of 2.1, 4.1, and 6.5 mJy beam$^{-1}$.}
\label{presenta}
\end{figure}

G26.47+0.02 (hereafter G26) is an LBV star candidate. This source presents a compact mid-infrared nebula, which closely resembles 
the ring nebulae around the LBV star candidates G79.49+0.26 and Wra 17-96 \citep{wachter10,clark03,egan02}. According 
to \citet{clark03}, adopting a distance of 6.5 kpc, 
the G26 central star has a luminosity of $10^{6}$ $L_{\odot}$ and a mass-loss rate of  $9 \times 10^{-5}$ \msol yr$^{-1}$,  
which indicated that it could be one of the most extreme stars in the Galaxy, similar to the known LBVs AG Car and AFGL 2298. 
Additionally, the authors suggested that G26 is photometrically variable.
Very recently \citet{naze11} performed an X-ray survey of Galactic LBV stars. The authors reported the detection of  G26 by Chandra 
and concluded that the X-ray emission is well explained by wind-wind collisions in a possible binary system. 

In Figure \ref{presenta} we present the LBV star candidate G26 in a color-composite image with the {\it Spitzer}-IRAC 8 $\mu$m 
emission (green), {\it Spitzer}-MIPSGAL emission at 24 $\mu$m (red), and the radio continuum emission at 20 cm 
(blue with white contours). 
We used the mosaicked image from GLIMPSE in the {\it Spitzer}-IRAC band at 8 $\mu$m, which
has an angular resolution of $\sim$1\farcs9 (see \citealt{fazio04} and \citealt{werner04}).
MIPSGAL is a survey of the same region as GLIMPSE, using the MIPS instrument (24 and 70 $\mu$m) on {\it Spitzer}.
The MIPSGAL resolution at 24 $\mu$m is 6\s. On the other hand, the radio continuum data at 20 cm with a FWHM synthesized 
beam of about 5\s~was extracted from the New GPS of the Multi-Array Galactic Plane Imaging Survey \citep{helfand06}.
It can be appreciated from Fig. 1 that the nebula surrounding the central star 
brightens at 24 $\mu$m and presents weak emission at 8 $\mu$m, a different case compared with the LBV star G24.73+0.69, 
which is very bright in both bands.
In addition, G26 has associated radio continuum emission, as was previously reported by \citet{clark03}. 

Bellow we study the interstellar medium around G26 and perform a multiwavelength analysis toward the central stellar object 
to unveil its nature.

\section{The environment of the LBV star candidate G26}

\subsection{HI absorption: distance estimate}

In this section we use HI data extracted from the VLA Galactic Plane
Survey (VGPS; \citealt{stil06}), which have an angular and spectral resolution of $\sim$1\m~and 1.56 \k, respectively.

\begin{figure}[h]
\centering
\includegraphics[width=8cm,angle=-90]{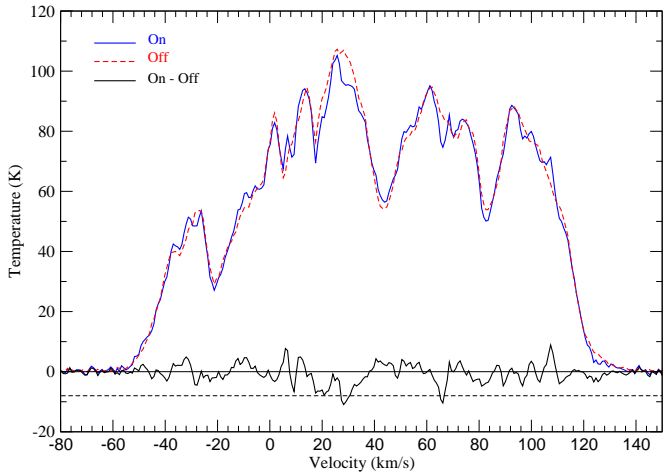}
\caption{HI spectra obtained toward G26.47+0.02 and its surroundings. The spectrum obtained toward the source
(the On position) is presented in blue, in red is presented the averaged HI emission
taken from four positions separated by approximately two beams from the source in the direction of the four galactic
cardinal points (the Off position), and the subtractions between them is presented in black. The 3$\sigma$
uncertainty of the subtraction is $\sim$8 K, which is shown with the dashed line.}
\label{hiabs}
\end{figure}

Taking into account that G26 presents radio continuum emission at 20 cm, it is possible to estimate its distance through a 
study of the HI absorption and the related molecular gas.
Figure \ref{hiabs} shows the HI spectra toward the source and surroundings.
The HI emission obtained over the source (the On position: a beam
over the radio maximum of the source) is presented in blue, in red is presented the average HI emission taken 
from four positions separated by approximately two beams from the source in direction of the four galactic cardinal points (
the Off position),
and the subtraction between them is presented in black, which has a 3$\sigma$ uncertainty of $\sim$8 K.
The figure shows that the last absorption feature appears
at v$_{\rm LSR}$ $\sim$ 66 \k. From the Galactic rotational model of \citet{fich89} this velocity implies 
the kinematic distances of either 4.3 kpc or 11.0 kpc. Taking into account that the tangent point (at v$_{\rm LSR}$ $\sim$ 119 \k) 
does not present any absorption, 
following \citet{kolpak03}, we favor the near kinematic distance. Considering that \citet{clark03} used 6.5 kpc as a maximum 
distance to the source, we conclude that 
G26 should be located between 4.3 and 6.5 kpc. In the following section, based on this result and by considering the presence 
of related molecular gas, the distance is constrained with better precision.

\subsection{Molecular gas}
\label{mol}

We use molecular data extracted from the Galactic Ring Survey (GRS). The GRS was performed by the Boston University and the
Five College Radio Astronomy Observatory. The survey maps the Galactic Ring in the \3 J=1--0 line
with an angular and spectral resolution of 46\s~and 0.2 \k, respectively (see \citealt{jackson06}).
The observations were performed in both position-switching and On-The-Fly mapping modes, achieving an
angular sampling of 22\s.

\begin{figure}[h]
\centering
\includegraphics[width=9cm]{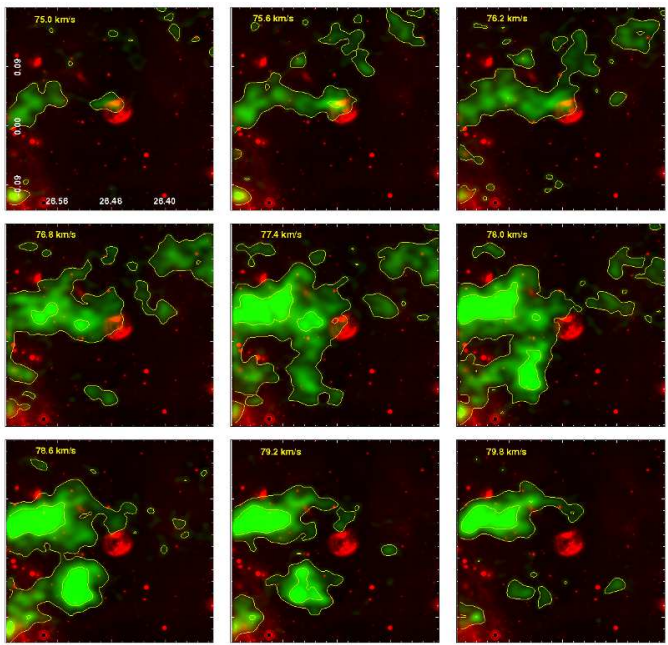}
\caption{Integrated velocity channel maps of the \3 J=1--0 emission (in green) every $\sim 0.6$ \k. The contour levels are 0.3 and 0.8 K \k.
Red is the 24 $\mu$m emission.}
\label{panel}
\end{figure}

\begin{figure}[h]
\centering
\includegraphics[width=8cm]{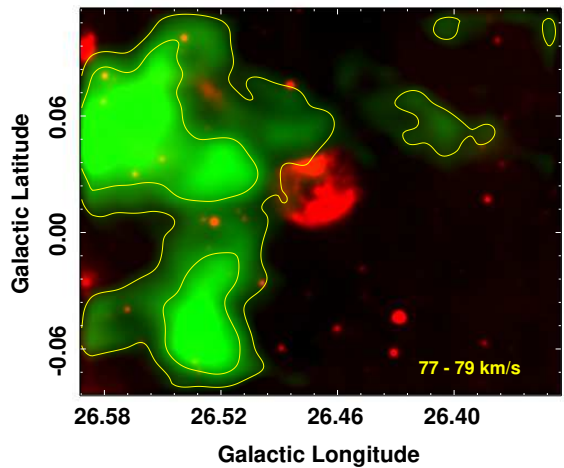}
\caption{\3 J=1--0 emission (in green) integrated between 77 and 79 \k. The contour levels are 1 and 2 K \k.
Red is the 24 $\mu$m emission.}
\label{integ77}
\end{figure}

\begin{figure}[h]
\centering
\includegraphics[width=8cm]{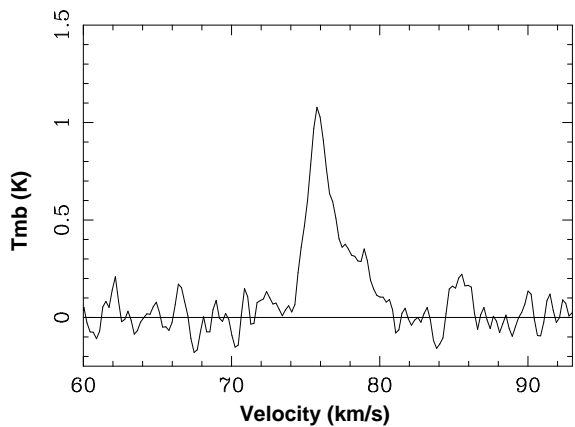}
\caption{\3 J=1--0 spectrum obtained toward the peak of the molecular feature that it is probably in contact with the northern 
brightest border of the infrared shell.}
\label{13cospectrum}
\end{figure}

In our search for molecular gas associated with G26, we analyzed the whole \3 J=1--0 data cube and found 
some interesting molecular structures likely related to G26 between 75 and 80 \ks (all velocities are given with respect to 
the local standard of rest).
Figure \ref{panel} displays, over the MIPS 24 $\mu$m emission (in red), the integrated velocity channel maps of the \3 J=1--0 
emission every $\sim$ 0.6 \ks (green with 
yellow contours), showing the kinematical and morphological structure of the molecular gas probably related to the G26 infrared shell. 
From the panels at 75.0 and 75.6 \ks a molecular clump can be seen, whose peak coincides with the northern brightest portion of 
the infrared shell, where
it presents an oblate shape, suggesting an interaction between them. This molecular clump is seen up to $\sim 79$ \k.  In the 
following panels the molecular gas appears to 
border the northeastern and eastern border of the 24 $\mu$m emission, forming an incomplete molecular shell with the lower 
level contour emission over the IR shell, 
likely produced by the effect of the strong stellar winds, as was found in the LBV star G24.73+0.69 \citep{albert11}. 
To better appreciate the molecular
gas bordering the north and northeastern portion of the 24 $\mu$m emission, we present in Fig. \ref{integ77} the \3 J=1--0 emission
integrated between 77 and 79 \k. Additionally, we show in Fig. \ref{13cospectrum} the \3 J=1--0 spectrum
obtained toward the peak of the molecular feature that it is probably in contact with the northern brightest border of the 
infrared shell (see panels at 75.0 and 75.6 \ks 
in Fig. \ref{panel}). 
Clearly, the spectrum is not symmetric and it presents a slight spectral shoulder or a less intense component at ``redshifted'' 
velocities. It could be evidence 
of turbulent motion in the gas, maybe produced by the strong winds of G26 (see e.g. \citealt{falgarone94}).

Taking into account the molecular study presented above, we suggest that the molecular gas is very likely related to G26, and 
hence we adopt the systemic  
velocity of 77.5 \ks for the source, deriving a distance of about 4.8 kpc. Hereafter we use this distance for G26.

From the HI and \3 J=1--0 spectra obtained towardxs the source we calculate the total hydrogen column density through 
N$_{\rm H}$ $=$ N(HI) + 2N(H$_{2}$), which is an 
important factor to compare with that obtained from the X-ray emission modeling (see Sect. \ref{x-ray}). From the HI data 
we obtain N(HI) $\sim$ 1.3 $\times 10^{22}$ cm$^{-2}$. On the other hand, by assuming that the \3 J=1--0 line is optically 
thin and using the typical LTE formula as used 
in \citet{albert11} with T$_{\rm ex} = 10$ K, we obtain N($^{13}$CO) $\sim 6 \times 10^{15}$ cm$^{-2}$. Using the relation 
N(H$_{2}$)/N($^{13}$CO) $\sim 5 \times 10^{5}$ \citep{simon01}, we obtain N(H$_{2}$) $\sim 3 \times 10^{21}$ cm$^{-2}$, 
yielding a total hydrogen column density of N$_{\rm H}$ $\sim 2 \times 10^{22}$ cm$^{-2}$.

To roughly estimate some physical parameters of the molecular clump associated with the northern maximum 
at 24 $\mu$m of the IR shell (see panels at 75.0 and 75.6 \k), we again assumed LTE and T$_{\rm ex} = 10$ K to derive a column 
density of N(H$_{2}$) $\sim 1.6 \times 10^{21}$ cm$^{-2}$ 
toward this structure. The integration was made between 75 and 80 \k. The molecular mass was estimated by performing the summation 
of this integrated emission in an elliptical area, centered at $l =$ 26\fdg479, $b =$ 0\fdg036 with major and minor axes 
of 55\s~and 40\s, respectively and an orientation angle of 0\d. 
We obtain a mass of M $\sim 125$ \msol, and assuming an ellipsoidal volume we derive a density of n$_{\rm H_{2}} \sim 600$ cm$^{-3}$, 
which suggests that the expansion of the infrared 
nebula is indeed encountering a relatively dense molecular clump toward the north.

\section{Multiwavelength analysis towards G26}

\begin{figure}[h]
\centering
\includegraphics[width=8cm]{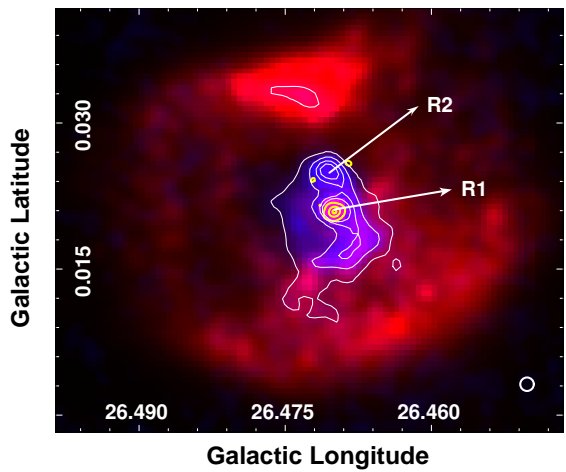}
\caption{Two-color composite image where the {\it Spitzer}-MIPSGAL emission at 24 $\mu$m is displayed in red and in blue appears 
the radio continuum emission 
at 20 cm with white contours with levels of 2.1, 4.1, 6.7, and 9.0 mJy beam$^{-1}$.  The yellow contours show qualitatively 
the X-ray emission obtained from Chandra observations (ObsID = 7493; see Sect. \ref{x-ray}). The beam of the radio continuum 
emission appears in the bottom right corner.} 
\label{radiofig}
\end{figure}

The LBV star candidate G26 brights in several wavelengths, which allows us to perform a complete study of this object. 
Figure \ref{radiofig} shows G26 
at 24 $\mu$m (red) with the radio continuum emission at 20 cm (in blue with white contours) and some X-ray contours shown 
in yellow, suggesting that all the emissions are related to the same object. Bellow we study the radio continuum, 
the millimeter continuum, and the X-ray emissions in sections \ref{radio}, \ref{mm}, and \ref{x-ray}, respectively.

\subsection{Radio continuum emission}
\label{radio}

The radio continuum data toward G26 at 1.4 and 4.8 GHz were extracted from the new GPS of the Multi-Array Galactic Plane Imaging 
Survey \citep{helfand06}. The emission at 1.4 GHz is presented in Fig. \ref{radiofig} (described above). Since the image at 4.8 GHz 
does not show significant morphological difference with that at 1.4 GHz, we do not show it here.

The dominant feature of the overall morphology is the strong 
contrast between the brighter (W) and much fainter (E) halves of the 
radio emission. The outermost boundary traces an almost elliptical shape with semi-axes of 24$^{\prime\prime}\times 18^{\prime\prime}$. 
Most of the radio emission is concentrated in an elongated band that runs  
approximately from north to southeast, which has two radio maximums. These maximums are cataloged as the radio sources GPRS 26.470+0.021 and 
GPRS 26.470+0.025 \citep{becker94}, R1 and R2 in Fig. \ref{radiofig}, respectively.
The brightest one (R1), and the only resolved with this data set, is coincident with the X-ray emission. It is 
centered at $l =$ 26\fdg469, $b =$ 0\fdg021 (18$^{\rm h} 39^{\rm m} 32^{\rm s}.2, -05^{\circ} 44^{\prime} 20^{\prime\prime}.7$, J2000).
The flux densities of R1 at both frequencies were derived by integrating the radio emission over a circular area of 11\s~in diameter defined from 
the 4.8 GHz image, in which the emission is much brighter. We obtained an integrated flux density of 0.025 Jy at 1.4 GHz and 
0.05 Jy at 4.8 GHz.  The other radio source (R2), a point-like one, is
located  on the northern border of the radio emission,  at $l =$ 26\fdg470, $b =$ 0\fdg024
(18$^{\rm h} 39^{\rm m} 31^{\rm s}.3, -05^{\circ} 44^{\prime} 12^{\prime\prime}.1$, J2000). Its peak has a flux density of 0.011 Jy at 1.4 GHz and 0.009 Jy
at 4.8 GHz. 
The errors in fluxes are about 20\% for both sources at each frequency. In both cases the contributions of the extended surrounded emission were subtracted. 
Thus, we estimate the radio spectral 
index $\alpha$ (S $\propto~ \nu ^{\alpha}$) between both frequencies, obtaining $\alpha \sim +0.57$ for R1, which is an expected value for thermal emission from the star, 
and  $\alpha \sim -0.26$ for R2, suggesting a non-thermal radio source.
On the other hand, to obtain the flux density of the extended radio emission, we subtracted the contribution of the two radio 
sources. We obtain a value of $\sim$0.12 Jy and 0.04 Jy at 1.4 and 4.8 GHz, respectively. In this case the error in the estimate of the flux densities is about 25\% and 
the intrinsic noise of each image was taken into account as well as the uncertainty in the choice of the integration boundaries. The radio spectral index of the extended emission 
turns out to be $\sim -0.9$, clearly non-thermal in nature. 

The observed non-thermal radio continuum should be due to synchrotron emission of relativistic electrons. The acceleration of the free electrons can be 
attributed to first-order Fermi acceleration in shocks within the stellar winds. These shocks can arise either from wind instabilities, or, in 
the case of massive binary systems, close to the contact discontinuity where the stellar winds of two stars collide (e.g. \citealt{dou00} and references therein).
It is known that Wolf-Rayet stars,  which lie on the same evolutionary path as the LBVs, frequently have binary companions, which suggests that it would be common
that LBV stars form binary systems  \citep{duncan02}. One possibility is that G26 is a binary system composed of R1 and R2, which are separated by $\sim$15\s~($\sim$0.3 pc or 
62 $\times 10^{3}$ A.U., assuming the distance of 4.8 kpc), and the non-thermal radio continuum emission arises from colliding winds. If this is the 
case, we are indeed looking at a very wide binary system (see e.g. \citealt{shaya11,caballero09}). Another scenario could be that R2 is not a companion star and/or it is just
a clump in the non-thermal radio continuum emission. In this case, the extended radio emission would be due to wind instabilities from R1, or colliding winds from R1 and 
an unseen stellar component.

\subsection{Near-infrared sources}
\label{nir}

\begin{figure}[h]
\centering
\includegraphics[width=8cm]{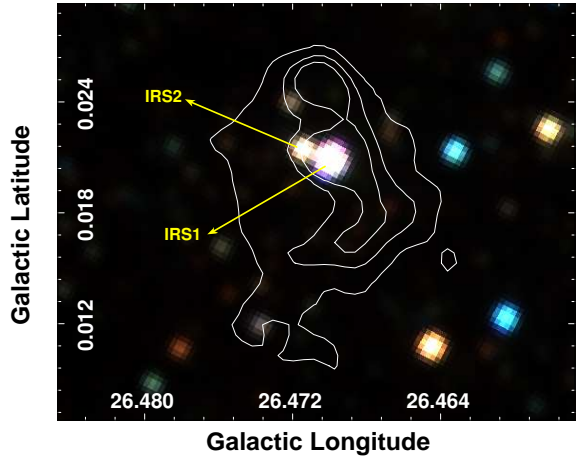}
\caption{2MASS {\it JHK} three-color image toward G26 ({\it J} in red, {\it H} in green, and {\it K$_{s}$} in blue). The white contours correspond to the radio continuum emission 
at 20 cm, whose levels are 2.1, 4.1, and 6.1 mJy beam$^{-1}$.}
\label{2mas}
\end{figure}

\begin{figure*}[h]
\centering
\includegraphics[totalheight=0.25\textheight,viewport=-30 0 490 700,clip,angle=-90]{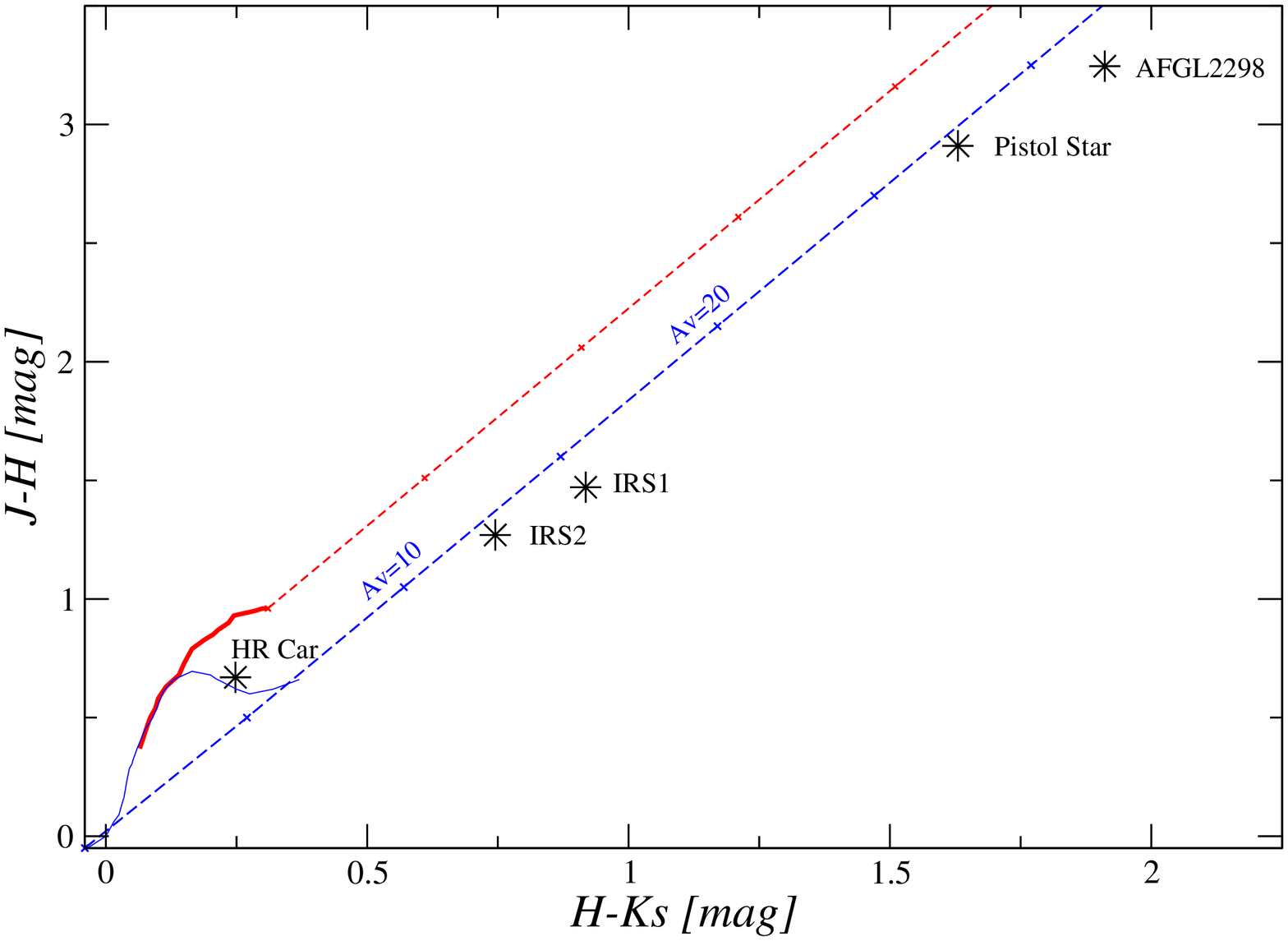}
\includegraphics[totalheight=0.265\textheight,viewport=-30 0 490 750,clip,angle=-90]{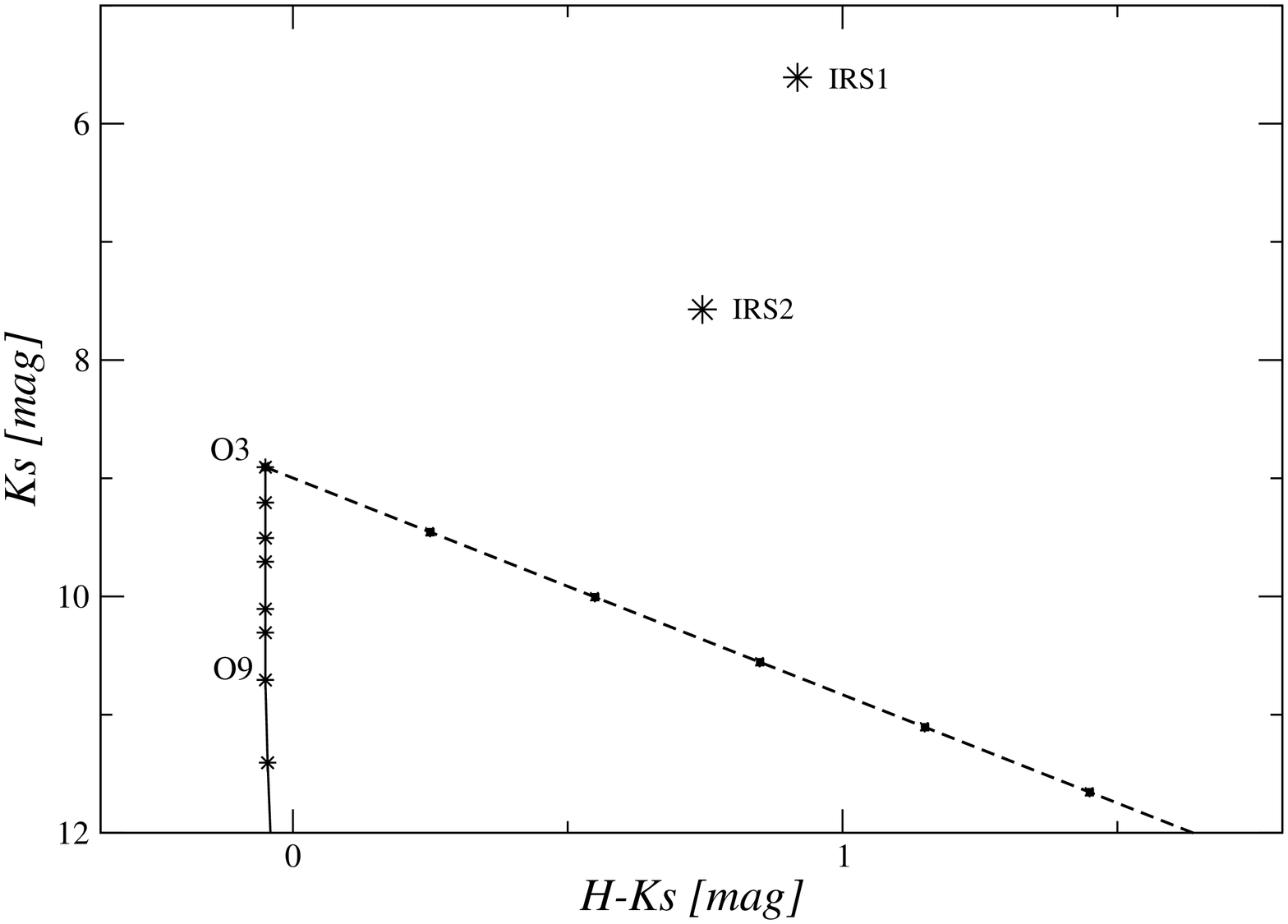}
\caption{Left: Color-color diagram showing the position of IRS1 and IRS2. For comparison we include some confirmed LBV stars.
The two solid curves represent the location of the main
sequence (thin line) and the giant stars (thicker line) derived from
\citet{bessell88}. The parallel dashed lines are reddening
vectors with the crosses placed at intervals corresponding to five magnitudes of visual extinction.
We assumed the interstellar reddening law of \citet{rieke85}. Right: Color-magnitude diagram of IRS1 and IRS2
The solid curve represents the position of
the main sequence at a distance of 4.8 kpc. The reddening vector for an
O3-type star, with the squares placed at intervals corresponding to five magnitudes
of visual extinction, is shown with a dashed line.}
\label{fot}
\end{figure*}

To look for a possible stellar companion of R1, we inspected the Two Micron All-Sky
Point Source Catalog (2MASS, \citealt{cutri03}) and extracted an image from the 2MASS Image Inventory Service. Figure \ref{2mas} shows a near-infrared (NIR) {\it JHK} three-color image 
toward G26 with the contours of the radio continuum emission overlaid. 
We show two infrared sources, IRS1 (2MASS J18393224-0544204) coinciding with the radio source R1, and IRS2 (2MASS J18393228-0544148), which is located $\sim5$\s~
northeastern from IRS1.  
In Table \ref{2masT} we present the 2MASS photometric data of IRS1 and IRS2. The errors in the magnitudes are included between parentheses. 
Both sources appear, in a typical ({\it J-H}) versus ({\it H-K$_{s}$}) color-color diagram (Fig. \ref{fot} left), as main-sequence stars with a similar visual absorption, suggesting that 
they could be located at the same distance. For comparison, we include some confirmed LBV stars in this diagram, such as HR Car, AFGL 2298, and the Pistol Star. 
Then, by assuming a distance of 4.8 kpc, we construct a {\it K$_{s}$} versus {\it H-K$_{s}$} color-magnitude diagram (Fig. \ref{fot} right), which shows that both sources are located 
considerably above the reddening track of an O3V-type star, remarking that they are giant stars, mainly IRS1. In a recent study of a stellar cluster located in the W33 complex, 
\citet{messineo11} presented a similar color-magnitude diagram in which blue and red super giants, WR and a cLBV star appear quite
above the reddening track of an O3V-type star, as in our case.  This rough NIR photometric study suggests
that IRS1 and IRS2 are giant stars probably located at the same distance, and hence they could be companions forming a binary system. However, we cannot discard the 
hypothesis presented above of the very wide binary system composed 
of the radio sources R1 and R2. The R2 source is associated with the IR source 2MASS J18393143-0544132, which is detected only in the {\it K$_{s}$} band ({\it K$_{s}$} = 13.079), 
and it is impossible to perform photometry.

\begin{table}[h]
\caption{2MASS photometric data of IRS1 and IRS2.}
\centering
\begin{tabular}{ccccc}
\hline\hline
Source & {\it J}  &  {\it H}   & {\it K$_{s}$} & Quality  \\
        & (mag)   &    (mag)   &   (mag)  &          \\
\hline
IRS1   & 7.997(0.021)    & 6.526(0.026)      & 5.608(0.017)    & AAA  \\
IRS2   & 9.587(0.126)    & 8.317(0.100)      & 7.572(0.110)    & BAB  \\ 
\hline
\end{tabular}
\label{2masT}
\end{table}

\subsection{Millimeter continuum emission}
\label{mm}

\begin{figure}[h]
\centering
\includegraphics[width=8cm]{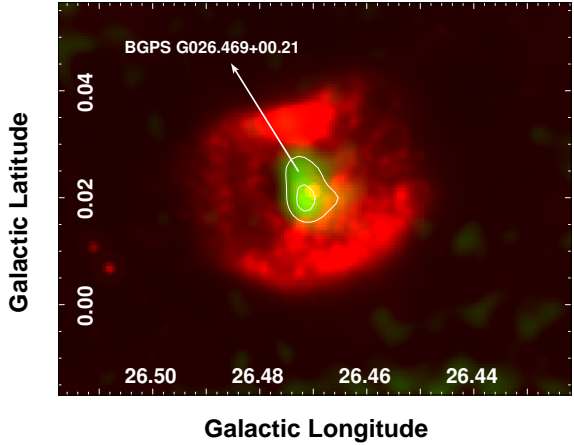}
\caption{Two-color composite image where the Spitzer-MIPSGAL emission at 24 μm is displayed in red and
the smoothed BGPS 1.1 mm continuum emission is presented in green with white contours with levels of 100 and 170 mJy beam$^{-1}$.}
\label{fig_bolo}
\end{figure}

Millimeter continuum emission from stars may result from thermal free-free, non-thermal synchrotron,
and thermal dust emission \citep{palla96}.
We used the Bolocam Galactic Plane Survey (BGPS) to study continuum emission at 1.1 mm toward G26.
The BGPS is a 1.1 mm continuum survey of the Galactic Plane made using Bolocam on the Caltech Submillimeter Observatory
with 33\s~FWHM effective resolution \citep{aguirre}.
We found a source from the BGPS located right upon the LBV star candidate,
namely G026.469+00.021, which is displayed in Fig. \ref{fig_bolo}. The positional coincidence
between G26 and the millimeter source points to a physical connection between them.

We first explore the possibility that the millimeter continuum emission originates only in the dust.
Following \citet{roso10}, we derive the mass of the emitting dust using
\begin{equation}
M=0.13D^{2}F_{\nu}\frac{e^{13/T_{d}}-1}{e^{13/20}-1}M_{\odot},
\label{eq1}
\end{equation} 
where $D$ is the distance in kpc, $F_{\nu}$ is the flux at 1.1 mm in Jy, and $T_{d}$ is the dust temperature in K.
This equation assumes an excitation temperature of 20 K. 
Taking $T_{d}=20$~K and a distance of 4.8 kpc, and using the total integrated flux 
of the source in the 1.1 mm band ($F_\nu=204$~mJy), we obtain a mass of dust of 0.6~\msol. 
It is unlikely that this dust component is linked to an ejection event of G26 for two reasons.
First, as noted by \citet{boyer10}, cold dust (T$<$100 K) around LBVs may correspond to pre-existing ISM dust swept up by stellar winds.
Second, the mass of dust is more than an order of magnitude larger than the mass estimated by \citet{clark03} ($\sim$0.019~\msol). 
Moreover, assuming a gas-to-dust ratio of 100, we obtain a gas mass of 60~\msol, which is greater 
than the mass of the ejection nebulae around other LBVs (see \citealt{smith06} and \citealt{clark09}). 
However, in a submillimeter study of $\eta$ Carinae, \citet{gomez06} estimated a mass of dust 
between 0.3 and 0.7 \msol, which may indicate that the gas-to-dust ratio in massive stars 
may differ from the canonical Galactic value.
We conclude that if the millimeter continuum emission originates in the dust, it is probably be pre-existing ISM dust swept up by stellar winds.

As we showed in Sect. \ref{radio}, the position of G26 coincides with the radio source R1, which has a positive spectral index.
This points to a thermal origin for the radio continuum emission. 
Therefore, we investigated the possibility that the millimeter emission originates in the ionized stellar wind 
by means of the thermal free-free mechanism. Then, we analyzed the spectral energy distribution (SED) in the radio and millimeter regimes, which relates 
the flux $F_{\nu}$ to the wind parameters. 
Assuming that G26 is located at 4.8 kpc, we scaled the stellar parameters derived by
\citet{clark03} using the proportionalities
introduced by \citet{hillier98}: $L\propto d^{2}$, $\dot{M}\propto d^{1.5}$,
$T_{{\rm eff}}\propto d^{0}$, where $L$ is the luminosity,
$\dot{M}$ is the mass-loss rate, and $T_{{\rm eff}}$ is the effective temperature.
For the luminosity, \citet{clark03} quoted $log(L/L_{\odot})\leq6.0$ for a distance
of 6.5 kpc, and this value drops to
$\sim$ 5.4 for a distance of 4.8 kpc. This makes G26 one of the faintest LBVs (see
the HR diagram of \citealt{clark09}) and a post-red-supergiant-phase candidate \citep{deJager98}. 
For the mass-loss rate, the value obtained by \citet{clark03}, i.e. $\dot{M}=9.5\times10^{-5}$~$M_{\odot}/{\rm yr}$, drops to $5.7\times10^{-5}$~$M_{\odot}/{\rm yr}$.
Finally, the temperature does not scale with the distance. 

After scaling the stellar parameters according to the new distance for G26, we use the following equation \citep{blomme11} to analyze the SED:
\begin{equation}
F_{\nu}[{\rm mJy}]=0.023\frac{1}{D^{2}}\left ( \frac{\dot{M}}{\mu v_{\omega}} \right )^{^{4/3}} \left ( g \gamma \bar{Z}^{2}\nu[{\rm GHz}] \right )^{^{2/3}},
\label{eq2}
\end{equation}
where D is the distance in kpc, $\dot{M}$ is the mass-loss rate in units of $10^{-6}$~\msol~yr, $\mu$ is the mean atomic mass, v$_{\omega}$ 
is the terminal wind velocity in units of 1000~\k, $\gamma$ is the electron-to-ion density number, 
$\bar{Z}=\frac{1+4\alpha_{{\rm He}}}{1+\alpha_{{\rm He}}}$ with 
$\alpha_{{\rm He}}$ the helium number abundance, and $g=9.77\left [ 1 +0.13\log_{10} \left ( \frac{T^{3/2}}{\bar{Z}\nu} \right ) \right ]$ 
is the Gaunt factor, with T the wind temperature in K.
The frequency dependence of the Gaunt factor is weak and the free-free emission from an ionized stellar wind 
can be approximated by $F_{\nu}\propto\nu^{0.6}$.

\begin{figure}[h]
\centering
\includegraphics[width=8cm]{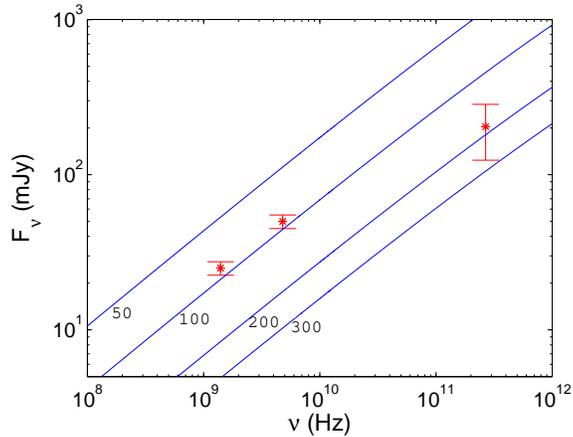}
\caption{Solid lines are the SED of an ionized stellar wind assuming different wind velocities (in \k).
The wind parameters used for the calculation are discussed in the text.
The dots are the 20 cm (1.4 GHz), 6 cm (4.8 GHz), and 1.1 mm (268 GHz) fluxes presented with their errors.}
\label{fig_SED}
\end{figure}

Assuming a helium abundance $\alpha_{{\rm He}}=0.1$, we obtain $\mu=1.3$ and $\bar{Z}=1.3$.   
We take the mass-loss rate from \citet{clark03}, scaled to a distance of 4.8 kpc,
i.e. $\dot{M}=5.7\times10^{-5}$~$M_{\odot}/{\rm yr}$, as shown above.
The wind temperature is set equal to 
the stellar temperature ($\sim$17,000 K). For a fully ionized H and He wind, we obtain $\gamma=1.1$.
In Fig. \ref{fig_SED} we plot the SEDs obtained from eq. \ref{eq2} 
for different terminal wind velocities (solid lines) and the fluxes in the radio bands at 20 cm (1.4 GHz) and 6 cm (4.8 GHz),
and the 1.1 mm band (268 GHz). 
Taking into account that the BGPS has lower angular resolution than the radio observations, the 1.1 mm flux may include contributions from both R1 and R2 radio sources. 
However, the expected flux for R2 at 1.1 mm should be $\sim$5 mJy (estimated from the spectral index of R2, see Sect. \ref{radio}), 
which is negligible compared to the total flux ($\sim$240 mJy). Accordingly, we assume that the millimeter emission originates in R1.  
From Fig. \ref{fig_SED}, we see that the fluxes from G26 depart from the $F_{\nu}\propto\nu^{0.6}$ dependence.
Considering the fluxes in the three bands, we obtain $\alpha\sim0.37$. 
Only the spectral index between 20 cm and 6 cm ($\sim$0.57, see Sect. \ref{radio}) approaches the expected value. However, we cannot discard 
a free-free origin for the millimeter continuum emission.

From this analysis, we conclude that both thermal free-free and dust emission (or a combination of them) are plausible mechanisms to explain the presence of BGPS G026.469+00.021 toward G26.

\subsection{X-ray emission}
\label{x-ray}

To study the physical connection of the X-ray emission detected from G26 with the observations obtained at other wavelengths presented above,  
we have re-analyzed Chandra observations which were first presented by \citet{naze11}. Our X-ray study improves, with a better fit, the study performed by these 
authors for this particular object.

\subsubsection{X-ray data}

The field of G26 was observed with the ACIS camera of Chandra on 2008 March 10 for 19.5 ks (ObsID = 7493). Chandra observations were calibrated using 
CIAO (version 4.1.2) and CALDB (version 3.2.2). To exclude strong background flares that eventually affect the observations, we extracted light 
curves of photons above 10 keV from the entire field-of-view of the camera, and excluded time intervals up to 3$\sigma$ to produce a GTI file. 

We excluded bad pixels from the analysis using the customary bad-pixel file provided by the Chandra X-ray Center for this particular observation. 
We have searched the data for background flares, which are known to affect Chandra data, by examining the lightcurve of the total count rate. 
An additional analysis was performed using the FTOOLS tasks. The spectra were analyzed and fitted within XSPEC v11.0.1 \citep{arnaud96}.

\subsubsection{X-ray images}

\begin{figure}[h]
\centering
\includegraphics[width=8cm]{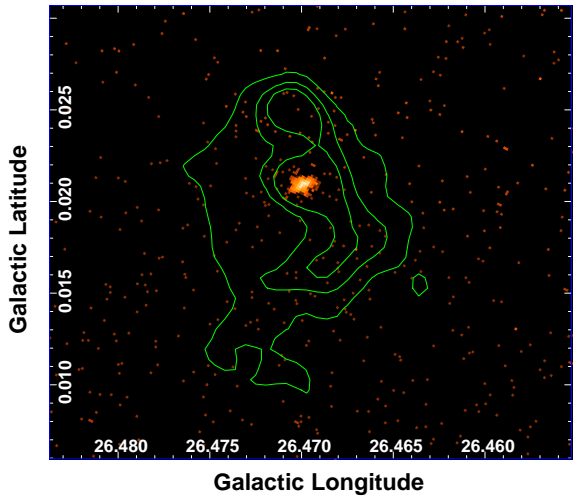}
\caption{ Chandra ACIS image, with a size of 0.4 $\times$ 0.4 arcmin,  
of G26 in the 1.0$-$5.0 keV energy band. The smoothed image was  
convolved with a Gaussian function of a kernel radio of 3$\times$3  
pixels ($\sim$ 1.5 arcsec). The green contours represent the radio continuum emission at 20 cm with levels of 2.1, 4.1, and 6.1 mJy beam$^{-1}$. }
\label{x-image}
\end{figure}

The sensitivity of the Chandra observations allowed us to detect a medium/hard X-ray point-like source at the 
geometrical center of the G26 infrared nebula. In the soft energy range (i.e. $<$1.2 keV) and above the 5.0 keV
the source is not detected. This source is located at $l = $ 26\fdg469, $b = $0\fdg020 (18\hh39\mm32\ss.219, $-$05\d44\m19\s.15, J2000), and has a signal-to-noise ratio of $\sim$ 7. 
In Fig. \ref{x-image}, we show an ACIS image in the energy band of 1-5.5 keV with radio contours at 1.4 GHz superimposed. As can be seen, the R1 peak is coincident with the 
point X-ray source detected by Chandra, and no X-ray emission is detected at the position of R2. 
To extract counts in the source region and compute net counts we used the dmextract tool. As a result, 400 counts were computed. 

Finally, to search for variability in the ACIS-I observation, we used  
the photon arrival times in the 1.0-5.0 keV band and the CIAO ``glvary''  
tool using the Gregory-Loredo algorithm \citep{greg92}. No hints of variability were detected during this observation.

\subsubsection{Spectral X-ray analysis}

\begin{figure}[h]
\centering
\includegraphics[width=6cm,angle=270]{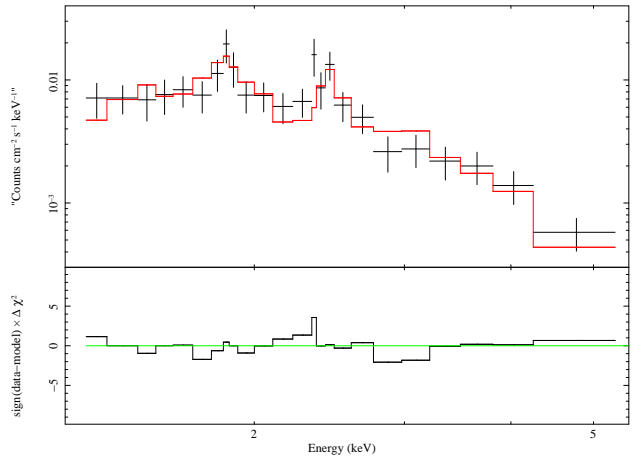}
\caption{Chandra X-ray spectrum of G26 in the 1.0-5.5 keV. The solid lines indicate the best-fit (VAPEC) model (see Table \ref{xtable}). 
Lower panel: Chi squared residual of the best-fit model.}
\label{x-spectra}
\end{figure}

We extracted the X-ray spectrum of G26 from a circular region with a radius of 4 arcsec using the CIAO tool specextract. The background spectrum was estimated from an annular region 
with radii of 4 and 10 arcsec. 
The spectrum is grouped with a minimum of 16 counts bin$^{-1}$ and the $\chi^{2}$ statistic is used. The errors quoted are 90$\%$.

The spectrum is shown in Fig. \ref{x-spectra} and it is well represented by a variable APEC model (VAPEC), which provides the best acceptable fit. 
The X-ray parameters for the best fit are given in Table \ref{xtable}.
The spectrum is dominated by atomic emission lines of Si and S, a temperature 
of $\sim$ 0.89 keV, and a high neutral hydrogen absorption column. 
These characteristics suggest that the X-ray emission has an optically thin thermal plasma origin.
Unfortunately, the poor quality of the data above 4 keV does not allow us to check for the FeXXV (6.4--6.7 keV) complex lines that could help us to constrain
the thermal and/or non-thermal emission at the hard part of the spectrum. For references about the detection of the FeXXV complex lines in massive stars 
see e.g. \citet{hyodo08}, \citet{albacete07}, and \citet{smith04}.
The obtained absorbing column density agrees well with that obtained from the CO and HI data (see Sect. \ref{mol}), confirming therefore the amount of material 
between us and the source. This fact supports the obtained radial velocity for the source and indirectly confirms the derived distance. 

\begin{table}
\caption{X--ray spectral parameters for G26}
\label{xtable}
\begin{center}
\begin{tabular}{llc}
\hline
\hline
Model && \\
Parameters && \\
\hline
PHABS && \\
$N_{\rm H}$ [cm$^{-2}$] && 3.46($\pm$0.43)$\times$10$^{22}$  \\
\hline
{\bf VAPEC} & & \\
\hline
$kT$ [keV]   & & 0.89$\pm$ 0.12\\
${\rm [Si/H]}$         &&  0.68$\pm$ 0.21  \\
${\rm [S/H]}$         & & 0.97$\pm$ 0.42 \\
Norm && 2.04($\pm$1.6)$\times$10$^{-3}$\\
\hline
$\chi^{2}_{\nu}$ / d.o.f. && 0.86 / 55  \\
\hline
Flux(1.0-2.5)[erg~cm$^{-2}$~s$^{-1}$] & & 1.69($\pm 0.02$)$\times$10$^{-12}$  \\
Flux(2.5-5.0)[erg~cm$^{-2}$~s$^{-1}$] & & 1.02($\pm 0.03$)$\times$10$^{-13}$  \\
\hline
Total Flux(1.0-5.0)[erg~cm$^{-2}$~s$^{-1}$] & & 1.79 ($\pm 0.06$)$\times$10$^{-12}$  \\
\hline
\end{tabular}
\end{center}
Normalization is defined as 10$^{-14}$/4$\pi$D$^2$$\times \int n_H\,n_e dV$,
where $D$ is distance in [cm], n$_{H}$ is the hydrogen density [cm
$^{-3}$], $n_e$ is the electron density [cm$^{-3}$], and $V$ is the volume [cm$^{3}$]. The
flux in the two energy ranges is absorption-corrected. Values in parentheses are the single parameter 90\% confidence interval.
The abundance parameter is given relative to the solar values of \citet{anders89}.
\end{table} 

We calculated the unabsorbed X-ray luminosities in the 1.5-5 keV band for a distance of 4.8 kpc, obtaining L$_{\rm X} \sim 4.7 \times 10^{33}$ erg s$^{-1}$,
and yielding log[L$_{\rm X}$/L$_{\rm BOL}$] $= -5.35 \pm 0.05$.
As mention above, the spectral X-ray analysis reveals that the emission has an optically thin thermal origin, which is dominant in the medium and hard 
energy range.
These characteristics seem to be indicating wind-wind collision shocks from a binary system of massive stars 
(e.g. \citealt{debecker05,pittard10}), reinforcing the hypothesis presented in Secs. \ref{radio} and \ref{nir}. 
This result does not seem to be common in LBVs. At present only very few X-ray emissions toward LBVs indicate
wind-wind collision shocks (see \citealt{naze11}). Indeed, as these authors point out, the LBVs as a class are clearly not bright X-ray emitters, and the few cases
of X-ray detections should have an extrinsic cause, such as binarity. If it is indeed common that LBVs possess companions like WRs do, following \citet{naze11}, there are basically 
two possibilities to explain the observed X-ray emission toward these objects: (i) the companion is quite close to the LBV, or (ii) the companion is relatively
distant from the LBV. In the first case, the intrinsic emission of the companion may be hidden by the strong absorption of the dense wind,
but then the conditions are favorable for an X-ray bright wind-wind collision. In the second case, no emission from a wind-wind collision
is expected due to wind dilution, but the intrinsic emission of the companion would be easily detectable, because the tenuous wind of the LBV cannot hide it anymore.  
High sensitivity observations are needed to explore these possibilities and the limit between them. If we consider this statement to be correct, 
the nature of the X-ray emission from G26 would discard the very wide binary system formed by R1 and R2 (see Sect. \ref{radio}).

Finally, it is interesting to mention that G26
was first detected by the ASCA telescope (source AX J183931-0544) by \citet{sugizaki01}. In that work the X-ray flux is a factor 2 greater
than the flux we obtained. However, this comparison should be taken with caution, no details about the model used to  
compute the ASCA flux and other parameters of the best-fit. Therefore, we cannot confirm a variable behavior of G26.

\section{Summary}

The luminous blue variable (LBV) stars are peculiar very massive stars. The study of these stellar objects and their surroundings is important for understanding
the evolution of massive stars and the impact that they produce in the interstellar medium.
We have performed a multiwavelenght study of the LBV star candidate G26.47+0.02 (G26) and its environment. The main results can be summarized as follows:

(1) G26 presents an infrared nebula that brights at 24 $\mu$m. We found that this nebula is partially surrounded by a molecular shell (seen in the $^{13}$CO J=1--0 line),
suggesting an interaction between the strong stellar winds from G26 and the molecular gas. From the HI absorption and the molecular gas study we conclude that 
G26 is located at a distance of $\sim 4.8$ kpc.

(2) Scaling the stellar parameters derived in previous works to the distance of 4.8 kpc, we found that G26 would be one of the faintest LBVs,
suggesting it to be a post-red-supergiant-phase candidate.

(3) The radio continuum emission at 6 and 20 cm shows two radio sources (R1 and R2) and extended emission around them within the G26 infrared nebula. Source R1 coincides with 
the central stellar object G26, and R2 is $\sim 15$\s~towards the north. From the radio spectral
index study we found that R1 presents thermal emission, while R2 and the extended emission are non-thermal. We suggest that the non-thermal radio emission 
might arise from the stellar wind instabilities, or, in the case of a massive binary system, from wind-wind collisions. In this scenario, R1 and R2
could be companions in a binary system.

(4) From the Two Micron All-Sky Point Source Catalog we found two IR sources with similar photomectrical characteristics, one coinciding with R1, and the other 
one $\sim 5$\s~toward the northeast. A rough {\it JHK} photometric study suggests that both sources are giant stars probably located at the same distance. 
We present it as another scenario for a possible binary system.

(5) We found that the 1.1 mm continuum Bolocam source G026.469+00.021 coincides with the G26 central source. We conclude that both thermal free-free and dust emission 
(or a combination of them) are plausible mechanisms to explain the millimeter continuum emission.

(6) Chandra observations reveal a medium/hard X-ray point-like source at the
geometrical center of G26, coinciding with the radio source R1. No X-ray emission is detected at the position of the radio source R2. 
No hints of variability were detected during the Chandra observation. Our results in the X-ray regime improve, through a better fit, the results obtained by \citet{naze11}. 
The spectral X-ray analysis reveals that the emission has an optically thin thermal origin, 
dominant in the medium and hard energy range. These characteristics seem to be indicating  wind-wind collision shocks from a binary system of massive stars.

Our main result points to the hypothesis that G26 could be a binary system. We give two possible scenarios for that: (i) a very wide binary system composed of the radio sources 
R1 and R2, and (ii) a binary system composed of the infrared sources IRS1 (coniciding with R1) and IRS2. With the present data we cannot exclude either of them. Binary systems appear 
to be common in WR stars and thus probably in LBVs. Therefore multiwavelength and long-term observations are needed to detect some possible variable behavior to confirm the 
binary nature of the system and to reliably determine its components.

\begin{acknowledgements}

We wish to thank the anonymous referee, whose comments and suggestions have helped to improve the paper.
S.P., J.A.C. and E.G. are members of the {\sl Carrera del 
investigador cient\'\i fico} of CONICET, Argentina. A.P. is a doctoral fellow of CONICET, 
Argentina. This work was partially supported by the following Argentinian grants: UBACyT 20020090300032 and 20020100100011, ANPCyT 2010-0008, PIP 112-200801-02166 (CONICET),
PICT 2007-00902 (ANPCyT), PICT 07-00848 BID 1728/OC-AR (ANPCyT), and PIP 2010-0078 (CONICET).
J.A.C. acknowledges support by DGI of the Spanish Ministerio de Educaci\'on y Ciencia under grants AYA2010-21782-C03-03, FEDER funds, Plan Andaluz de Investigaci\'on 
Desarrollo e Innovaci\'on (PAIDI) of Junta de Andaluc\'\i a as research group FQM-322 and the excellence fund FQM-5418.

\end{acknowledgements}

\bibliographystyle{aa}  
\bibliography{biblio}
\IfFileExists{\jobname.bbl}{}
{\typeout{}
\typeout{****************************************************}
\typeout{****************************************************}
\typeout{** Please run "bibtex \jobname" to optain}
\typeout{** the bibliography and then re-run LaTeX}
\typeout{** twice to fix the references!}
\typeout{****************************************************}
\typeout{****************************************************}
\typeout{}
}

\label{lastpage}
\end{document}